\documentclass[times]{smrauth}

\usepackage{booktabs}
\usepackage{cleveref}
\usepackage{dirtytalk}
\usepackage{natbib}
\usepackage{subfig}
\usepackage{float}
\usepackage{longtable}
\begin{document}

\runningheads{K. Al-Sabbagh, L. Gren}{}
\author{Khaled Al-Sabbagh\affil{1}\corrauth{} and Lucas Gren\affil{2}}

\address{
\affilnum{1}Chalmers University of Technology and The University of Gothenburg, The Department of Applied IT,
 Gothenburg, Sweden 412--96 \break
\affilnum{2}Chalmers University of Technology and The University of Gothenburg, The Department of Computer Science and Engineering, Gothenburg, Sweden 412--92 
}

\corraddr{Chalmers University of Technology and The University of Gothenburg, Gothenburg, Sweden 412--96. E-mail: khaals@chalmers.se}

\title{The Connections Between Group Maturity, Software Development Velocity and Planning Effectiveness}

\begin{abstract}

Empirical evidence regarding the connection between group development (maturity) and the success of software development teams is lacking. The purpose of this research is to gain a qualitative and quantitative understanding of how velocity and planning effectiveness of software teams connect to a group development model. The Group Development Questionnaire (GDQ) was given to 19 software developers from four work groups in order to assess their group development maturity. The work groups' responses to the survey were checked for correlation with development velocity and planning effectiveness. Semi-structured interviews were conducted with 16 individuals from the same four work groups to explore issues about their group maturity and to validate the responses of the interviewees in the GDQ. The measurement of the fourth stage of group development had a strong association with the planning effectiveness measurement, which means that a team with less issues in the fourth phase of group development is more effective in adhering to its plans. On the other hand, group development and velocity showed no significant convergent validity. We conclude that the dynamics within software development teams might correlate to their ability to deliver the expected outcome as planned, but not to their ability to develop tasks faster.

\end{abstract}


\maketitle
\section{Introduction}
Groups, like humans, move through successive phases; they tend to advance and regress \cite{tuckman1965}. A group is sometimes defined as three or more  members that interact with each other to perform a number of tasks and achieve a set of common goals \cite{grupp}. A team, on the other hand, has developed both the goals and the means to achieve these tasks effectively \cite{validationstudy}. The emphasis on the importance of arranging work in a group-form emerged, in part, from the growing awareness of the role of groups in facilitating or blocking individual and organizational effectiveness, and more work can be achieved in well-functioning teams than dividing work to individuals only \cite{hinsz1997emerging}. As a result, organizations are counting on teams as the main asset for accomplishing goals \cite{facilitating}. 

Group development can be defined as the process in which a group navigates a number of stages until it becomes a mature team. Consequently, the term ``group maturity'' refers to the level of development a group has acquired over the course of its lifespan. Wheelan et al. \cite{validationstudy} reported that 83\% of teams that were assessed in a study were found to be work groups without effective means to reach their common goals. A team, therefore, is here defined as one that has successfully navigated the earlier stages of group development and has emerged as a mature, high performing unit capable of achieving common goals \cite{wheelan2005}.

The work of Susan Wheelan on group development research helped determine the common threads among group development models and postulate the basis for the Integrated Model of Group Development (IMGD). In this model, a group is believed to go through five successive stages of development, namely ``Safety and Inclusion,'' ``Counter-dependency and Fight,'' ``Trust and Structure,'' ``Productivity and Work,'' and ``Termination.'' The importance of this model lies in the fact that it proposes a statistically validated instrument that measures the maturity of a given group at a given time, called the Group Development Questionnaire (GDQ). The instrument, developed by Susan Wheelan in 1993, contains four sub-scales bases on the stages from her IMGD. Each sub-scale contains 15 items which measure the amount of energy a group is spending on the corresponding stage of IMGD. A comprehensive validation study on the GDQ, performed by \cite{wheelan1996}, revealed reliability scores for scales one through four to be 0.74, 0.87, 0.69, and 0.82 respectively, which indicate a good overall reliability of the GDQ items. In this study, we used the IMGD model as the theoretical framework for understanding the group dynamics of the participating work groups and the GDQ was used to assess their group maturity. 

While team performance is defined as the extent to which a team is able to meet cost, time, and quality objectives, a differentiation between two variables, effectiveness and efficiency, needs to be made in order to gain insights into the actual performance of software teams. Effectiveness refers to the team's adherence to the predetermined quality of a product \cite{hoegl2001teamwork}. In a software context, effectiveness could be the robustness or reliability of functionality in software. Efficiency, on the other hand, is evaluated in terms of team's commitment to schedules \cite{hoegl2001teamwork}, like launching software on the target date and within budget. Therefore, effectiveness reflects a comparison of actual versus intended outcomes, whereas efficiency ratings are based on a comparison of actual versus intended input \cite{hoegl2001teamwork}. 

The performance of software teams are sometimes stated to essentially be measured using two extremes: objective and subjective \cite{ong2005team}. The subjective approach relies on the perception of key stakeholders (e.g., the customer) on the performance of a given team whereas the objective approach relies on a quantitative assessment of team performance \cite{ong2005team}. One way to measure the latter is to look at the team's adherence to schedule, just like our previous definition of efficiency, which are both only in relation to plans and not the customer value. In software teams that adopt scrum in their development, planning occurs on a sprint level, where all sizes of completed work items are collected at the end of a sprint to determine the velocity of the team \cite{agarwal2012tracking}. The value of the completed work is only recognized when the work gets accepted by the product owner at the end of the sprint. In other words, no points are given for any work done until it gets accepted. Based on this, we used the Schedule Performance Indicator (SPI) \cite{albero2014understanding} to measure the effectiveness of the scrum teams in planning their stories and delivering the expected outcome. In this current research, we use the term ``planning effectiveness'' to describe the teams' ability to deliver the planned work as expected in relation to customer value represented by the product owner. Also, we measured the velocity of the work groups in accomplishing their scrum tasks, at the end of a given sprint, by calculating the number of hours spent on them. As a result, the velocity measurement used in this research reflects the teams' efficiency in accomplishing scrum tasks while planning effectiveness reflects their ability to estimate and deliver, within each sprint, the expected outcome. 

Due to the overlap and, therefore, some confusion between the constructs of performance, productivity, effectiveness, efficiency, planning, quality, etc.\ we reduced our study to only comprising Schedule Performance Indicator (SPI) with story points as a measurement of effectiveness, and mean velocity of scrum tasks as a measurement of efficiency. 

It must be noted here that we do not claim that mean team velocity and planning effectiveness are by any means a complete measurement of performance. However, we believe the related work presented in Section 3 provides us with good reasons to believe that they are, at least, key factors in software development performance. 

Several studies that used group development psychology as a theoretical framework have been conducted to examine the effect of group maturity on the productivity of teams in different contexts \cite{school,intensivecare,facilitating}. This highlights the usefulness and versatility of understanding groups from this perspective. However, empirical evidence regarding the influence of group maturity on the success of software engineering work groups with innovative tasks is lacking. In fact, studies demonstrating a link between teamwork in the field of agile software engineering has just begun~\cite{grenjss2}, and the agile approach do imply more focus on teams, which calls for even more focus on such empirical studies. To the best of our knowledge, only one study investigates the link between group maturity and agility of software teams in more detail~\cite{grenjss2}. In the same authors' earlier conference publication~\cite{maturityAgility} they suggested the use of velocity as a factor to further validate their findings since the tool used in their study, Sidky's \cite{sidky}, is not thoroughly validated~\cite{grenjss}, which means that it might not even measure agility. Therefore, our research investigates the correlation between group maturity and velocity to address this gap, but also adds the aspect of planning effectiveness. Planning effectiveness was added since we also wanted to investigate another essential mechanism of software performance that is more dependent on the group dynamics than velocity is, i.e., developing features fast might not be useful if they do not add costumer value.  

\section{Background}\label{ch:related_work}

\subsection{Group Development}

Group development research began in the 1930s with the work of Lewin on group climate and conflict between groups. The study of the behavior of small groups was launched with the establishment of a research center for group dynamics in 1946 \cite{verbal}. 

Bion \cite{verbal} described the effect of emotional states on group development. The results revealed that two levels of activity are found in groups. One level is geared towards the accomplishment of tasks, known as \emph{work group}, whereas the other, known as \emph{basic assumption group}, interferes with tasks achievement. Dependency, fight-flight, and pairing were identified as the emotional states that deviate a group from its work task, i.e., the basic assumption group, but these are crucial for group cohesion. These are not necessarily sequential and they can occur at anytime during the life of a group. Bion's theory was further expanded by Slater \cite{verbal}, who postulated that the themes that affect group development are the relationship of members to its leader, its need for order, and its wish for immortality.

Likewise, the influence of emotional states on group development was recognized by a plethora of literature such as members experiencing negative emotions will influence team performance regardless of the team's level of integration \cite{rippleeffect}. The study also showed no significant difference between the effect of positive versus negative valence on emotional contagion, i.e., both negative and positive emotions would result in a contagion of moods among group members with varying degrees. 

An integrative theory of linear and cyclic models was first introduced in 1964 \cite{wheelan2005}. The theory postulates the existence of four primary elements in group development. \emph{Acceptance}, which focuses on the creation of trust and the reduction of anxiety, and the growth of self-confidence among members of the group. \emph{Data-flow}, involves the ability of a group to make decisions as a result of communicated feelings and data across its members. \emph{Goal Information}, relates to the group's productivity as evidenced by their ability to perform problem solving and decision-making. The final element is referred to as \emph{Control}, the degree by which members of the group are recognized as interdependent and organized \cite{wheelan2005}. 

A major comprehensive analysis of various group development models was conducted by \cite{tuckman1965}. In this analysis, 50 articles on group development were reviewed based on a classification system of three elements: 1) setting (such as a laboratory group, natural group, or therapy group) 2) task or social focus 3) stage of development. The result of this analysis was a conceptual model comprising of four stages of group development in which each stage has a social realm and a task realm. The four stages proposed by Tuckman are: \emph{forming} which is categorized by high dependency, orientation, and testing; \emph{storming} during which resistance to both tasks and the influence of a group is apparent; \emph{norming} in which opinions are more freely expressed; and \emph{performing} in which a focus is on tasks' accomplishment after structural issues in a group are resolved. A review of this model was made by Tuckman in which he added a fifth stage of \emph{adjourning} \cite{tuckman1965, tuckman1977}. This was following a review made by \cite{mills} on his four-stages model suggesting adding a separation and a conclusion stage. Tuckman's theory gained empirical support by many researchers \cite{wheelan2005}. 

In the review of a number of the studies that did not support the group development stage theory, \cite{cissna1984} pinpointed a number of erroneous approaches in the methods adopted in these studies consequently citing that ``every group is like all groups in some respect, and like no groups in other respect'' \cite{cissna1984}. Moreover, there is ample evidence in the body of literature which support the theory of stages in group development \cite{wheelan2005}. While these models share the same view that groups face a basic set of developmental changes over time, the differences persist in the recognition and labeling of each stage and their sub-components in the group development \cite{wheelan2005}. 

\paragraph{The Integrated Model of Group Development (IMGD)}
The IMGD was theorized after consolidating previous theories which proposed a unified group development model for all group types \cite{wheelan1994}. The overall goal of group development was set to establish an organized unit of members capable of working effectively to achieve specific goals. What follows is a description of the five stages, found in the IMGD, which describe the behavioral pattern of any group type \cite{wheelan1994}. 

\paragraph{Stage One}
The first stage is a period of \emph{Dependency and Inclusion}, where members tend to show significant dependency on the leader in resolving new issues. At this stage, members spend a significant amount of energy to achieve a feeling of safety and inclusion in their group. As a result, members become leader-focused, in a sense that she will provide protection for the members. Members are indulged in an exploratory phase for the sake of identifying their roles, rules, and the structure within the group. Their exploration is characterized by being tentative and overly polite since they fear being rejected \cite{wheelan2005}. 

\paragraph{Stage Two}
The second stage is referred to as \emph{Counterdependency and Fight}. At this stage, members feel freer to express conflict between each other or among members and leaders since some needs for safety have been achieved in the previous stage. The group tries to free itself from being leader focused, and tends to fight about the group's goals. Coser explained that conflict is an important part for the development of cohesion, as it provides the opportunity for setting the psychological boundaries, which facilitate the establishment of goals, shared values, and structure \cite{verbal}. The occurrence of conflict is a result of the members' attempt to reach a unified direction out of the many divergent viewpoints. The rise of coalitions between members who share similar values and ideas is very much apparent \cite{wheelan2005}.   

\paragraph{Stage Three}
After navigating the inevitable stage of conflicts, communication becomes more open and members' trust and cooperation increase. Feedback and information sharing increase rather than being kept as a way to gain power.\ The aforementioned characteristics consolidate a more solid and positive relationship between members, which allow the group to carry out more mature negotiations about their goals and procedures. The group is at a stage where it is designing and preparing itself to start working effectively. Although work occurs in all the stages of group development, the group's focus on structure and goals at this stage significantly increases the group's capacity to work more productively \cite{wheelan2005}. 

\paragraph{Stage Four}
As soon as the goals and structure of the group are set from the previous stage, the group's focus is diverged into getting the work done well at the same time as the group cohesion is maintained, and remains cohesive while engaging in task-related conflict. It is in stage four that the group can start self-organizing and the leader can step back and be an expert member of the team instead of helping directing the work \cite{wheelan2005}.
 
\paragraph{Stage Five}
Most groups, temporary or continuously, experience an ending point at some point in the course of their lives. At the ending point, functional teams tend to give feedback about each other \cite{wheelan2005}. It has been reported that this type of processing is important for individual members since it enhances their ability to work effectively in the future. Impending termination of a group alters its structure and is likely to result in the group's regression to earlier stages of the group development \cite{wheelan2005}. 
\subsection{Tools for Measuring Group Development}
Various self-reporting instruments have been developed in the last few decades to aid team building and highlight the importance of group development. In order to decide which instrument to use for measuring the maturity of the work groups, we reviewed some of the tools and investigated if they were statistically tested for validity and reliability. Below are some of the tools that we came across in our literature review:

\begin{itemize}
\item The Team Development Inventory (TDI)
\item The Group Development Stage Analysis
\item The Group Attitude Scales
\item and the Group Development Questionnaire (GDQ)
\end{itemize}

Our investigation showed that the GDQ has been studied thoroughly relative to validity and reliability \cite{wheelan1996}, which makes it an appropriate choice for measuring the maturity level of work groups. 

\paragraph{Group Development Questionnaire (GDQ)}

Based on the IMGD, the GDQ was developed after being subjected to a number of statistical tests for reliability and validity \cite{wheelan1996}. The 60-item instrument contains a total of four scales. Each scale contains fifteen items, which corresponds to a single stage in the IMGD. For copyright reasons, only three items from each scale are presented (see Table \ref{GDQ}). The instrument does not assess the termination stage since it is meant for use with existing groups only. Items on scale I measure the amount of energy a group is spending in dealing with issues of inclusion and dependency. Items on scale II seek to measure the amount of group focus on issues of counter-dependency and conflict. The group's current level of trust and structure is measured by scale III, which corresponds to stage three in the group development model whereas the group's maturity on the ``work and productivity'' is measured by scale IV \cite{wheelan1994}.

Internal consistency tests for each fifteen-item scale were performed to ensure that all items within each scale were consistent \cite{wheelan1996}. Furthermore, the instrument was correlated with the Group Attitude Scale to establish concurrent validity \cite{evans1986}. The results indicated a significant concurrent validity between the two measures. Moreover, criterion-related validity was investigated. Results showed that groups who ranked high on productivity had significantly lower scores on the first and second scales of the GDQ. Similarly, groups that ranked high on productivity had significantly scored high scores on the third and fourth GDQ scales \cite{wheelan1996}. 

\subsection{Soft Factors Affecting Software Team Performance}
Purna et al. \cite{purna2011soft} classified the factors that influence the performance of software teams into technical, non-technical (soft), organizational, and environmental. These
factors are interconnected with each other and together they contribute to the overall performance of a software development team. Since this study focuses on investigating the relationship between group maturity and some aspects of software performance, only soft factors were considered. Below are some of these factors that were shown to influence the performance of software development teams, as we came across in our literature review.

\paragraph{Team diversity}

Although team diversity stems from a myriad of reasons such as education, experience, ethnicity, culture, skills, age, gender, etc, The results from a study conducted by \cite{liang2007effect} showed that knowledge diversity in teams had positively influenced team performance whereas value diversity had a negative influence on teams performance \cite{liang2007effect}. 

\paragraph{Team member competencies and characteristics}

Competencies can be classified into two categories: technical and personal competencies \cite{asproni2004motivation}. Asproni \cite{asproni2004motivation} explained that personal competencies can sometimes outweigh the technical in their influence on team performance. For example, a team of junior programmers with high personal competencies can perform better than a team of senior software developers. Similarly, another study conducted by \cite{huckman2009team} in cooperation with a software development company in India confirmed that having team members who have previously worked with each other has a positive influence on team performance regardless of the years of experience of team members.

\paragraph{Conflicts in team}

In a study conducted by \cite{sawyer2001effects} on 40 software development teams found that team members’ characteristics and the intra-group conflicts explained half of the variance between good and bad performing teams. The results concluded that intra-group conflicts have a negative influence on the performance of software teams. Similarly, it was also shown by \cite{gren2017links} that some of the agile practices are negatively affected by interpersonal conflict. 

The IMGD model characterized a productive work group as one that has navigated the earlier stages of group development and has become more focused on building trust and structure, and work and productivity. As the IMGD describes some of the behavioral aspects manifested by groups in all the stages of group development, these aspects show similarity with some of the soft factors described above. For example, \cite{wheelan1994} described stage two in group development as a period of fight and counter-dependency where conflicts between members are prominent, which negatively affects the group performance. Likewise, Sawyer \cite{sawyer2001effects} suggested that conflict in teams is a significant factor that yields to a deterioration in team performance. Moreover, Wheelan \cite{wheelan2005} described a stage three group as one whose members communicate more openly, cooperate more effectively, and share information and feedback. Similarly, \cite{anderson1998measuring} proposed that a clarity on the team's goals, a safe working environment that supports idea sharing and the participation of individuals are key factors in positively affecting the performance of software teams.   

\subsection{Software process improvement models}
Over the years, the high rate of software failures has been a challenge confronted by many software organizations \cite{integratingimprovement}. In 2001, a survey conducted with over 8000 U.S software projects, a schedule overrun of 120\% and a project cancellation rate of 25\% were reported \cite{integratingimprovement}. The problem with most of these software projects is that they are run on an ad-hoc structure, where poor planning and high defect rate are common. Improving the outcomes of these projects requires an effort on three different levels: organization, team, and people \cite{TSPinGSD}. For each level, different models were developed to ensure continuous improvement and quality.
Models such as CMMI (Capability Maturity Model Improvement) operate on the organizational level, which is based on the premise that organizations are continuously looking for ways to evaluate and improve their current processes in the guide of achieving better quality. 
On the other hand, models such as TSP (Team Software Process) work on the team level, providing sound software engineering guidelines for engineers to create and maintain self-steering software teams. The model provides a framework for groups working on software-intensive projects to organize and manage their work via an iterative cycle of eight phases: launch, strategy, planning, requirements, high level design, implementation, integration and testing, and post-mortem \cite{TSPinGSD}. In all iterations, the launch stage is conducted to clarify goals and assign roles for group members. Members of work groups conduct regular meetings, usually weekly, to share data about their work including goals achieved, risks that have developed, and issues that have emerged. The openness in sharing these data promotes for an atmosphere of trust and structure in the work group, where members are encouraged to report, listen, and contribute in planning their work \cite{TSPRevolution}. The competitive advantage of implementing TSP was reported by several global organizations. For example, in a TSP project implemented by Hill Air Force Base, a U.S. government organization rated at CMM Level 5, a productivity improvement of 123\% and an average reduction of 20\% in the test time of the project schedule were reported \cite{humphrey2000personal}. Another example is in an avionics project carried out by Boeing where 94\% of system test time was reduced by the implementation of TSP resulting in substantial improvements in the project schedule and allowing Boeing to deliver a high-quality product ahead of schedule. 

 Lastly, models such as PSP (Personal Software Process), which is a prerequisite for TSP, deal with people. This model provides methods that allow individual engineers to improve their planning and reduce product defect rates. By utilizing PSP, practitioners learn how to manage and evaluate the quality of their work. The PSP model provides a set of advantages that improves the performance of software engineers. The results from a study conducted by \cite{humphrey2000guest} showed that by adopting PSP, engineers could overcome resistance to transition when introduced to new technology. Also, the model teaches engineers a wide variety of skills ranging from requirements and system design to testings and deployment. 

While all of these software improvement process models aim at improving software quality, they disregard the psychological element associated with changes in group dynamics over time within work groups and its influence on building mature teams. The TSP model was founded on the premise that building mature teams who are capable of cooperating tasks and working towards shared goals improve their planning effectiveness and work quality. Likewise, the IMGD model suggests that mature teams, who are at stage four of group development, are highly effective and deliver high quality products in a timely manner. Both models promotes trust and cooperativeness as a vehicle for teams to become more effective. However, unlike TSP, which is tailored specifically to software engineering teams, the IMGD model accounts for the group development stages and acknowledges its influence on building mature teams.  

\begin{table}

  \caption{An excerpt of the items contained in each Group Development Questionnaire (GDQ) scale} \label{GDQ} 
  
  \centering
  \begin{tabular}{|c|l|}
  \hline
     \footnotesize \textbf{Scale} & \footnotesize \textbf{Sample Items} \\ \hline
     \footnotesize GDQI & \footnotesize Members tend to go along with whatever the leader suggests \\
      &  \footnotesize There is very little conflict expressed in the group. \\
      &  \footnotesize We have not discussed our goals very much. \\
      
     \footnotesize GDQII & \footnotesize People seem to have very different views about how things should be done in this group \\
     & \footnotesize Members challenge the leader's ideas\\
     & \footnotesize There is quite a bit of tension in the group at this time \\
     \footnotesize GDQIII & \footnotesize The group is spending its time planning how it will get its work done \\
     & \footnotesize We can rely on each other. We work as a team. \\
     & \footnotesize The group is able to form subgroups, or subcommittees, to work on specific tasks. \\
     \footnotesize GDQIV & \footnotesize The group gets, gives, and uses feedback about its effectiveness and productivity  \\
     & \footnotesize The group acts on its decisions \\
     & \footnotesize The group encourages high performance and quality work \\ \hline
  \end{tabular}
     
\end{table}

\section{Related Work}
\paragraph{Application of IMGD in Different Contexts}
Several studies, adopting the IMGD as a theoretical framework, have been conducted to examine the effect of group maturity using GDQ on the productivity of teams in different contexts, highlighting the usefulness and versatility of this tool. One study looked at the learning outcomes of students in schools as measured by math, reading, and achievement ranks and the maturity level of school administrators as measured by GDQ. The study concluded a significant relationship between the functioning of faculty group and students' learning outcomes \cite{school}. Similarly, another study investigated the relationship between the level of teamwork in the Intensive Care Unit (ICU) and the patients' outcome. Data were analyzed by correlating the ICU mortality rate (patients' risk of dying in the hospital using a mortality prediction system) and stage of group development of 394 staff members in the participating 17 ICU in nine hospitals. A significant correlation was identified between a unit's stage of group development and that unit's mortality rate \cite{intensivecare}. As the staff perception of their level of group development increased, mortality rate in their unit decreased, i.e., the higher the level of group development a group is, the fewer deaths occurred. A third study used the GDQ to plan an appropriate intervention to improve the effectiveness of three work groups in semi-governmental organizations. In this study, the group development scores of the three groups on the four GDQ scales were determined, an appropriate intervention to improve the teams' effectiveness was devised, and a three-months follow-up plan was set to determine whether significant positive changes had occurred. The intervention revolved around the issues revealed from the GDQ data. For example, member discussion was encouraged to focus on the importance of hearing opinions from all team members, reducing the dominance of the leader without creating a hostile environment, etc. Paired samples tests were employed to determine whether the intervention resulted in a positive significance on the fourth GDQ scale and effectiveness ratio within each group from pre to post tests \cite{facilitating}.

These various models suggest that interactions within a group display predictable patterns and that human interactions affect work performance within a group. These models have been the result of mainly observation of groups functioning in different settings (laboratory group, natural group, therapy group, etc.).  The culmination of these models helped Susan Wheelan formulate the IMGD (Integrated Model for Group Development) which, unlike many other models, developed an instrument, the GDQ, to capture data on how groups behave and progress relative to stages of group development.

\subsection{Software Team Performance Measurement}

Ong et al. \cite{ong2005team} identified two approaches in which the performance of software development teams can be measured: objective and perceptual or subjective. The first approach includes measuring function points, object points, use case points, kilo lines of code, and defect rate. Sawyer \cite{sawyer2001effects} explained that perceptual measures, such as quality of the product and satisfaction with the product should be taken from external stakeholders in order to account for self-bias. The perceptual or subjective approach relies on the group's perception of their team performance and is based on items such as \emph{our group is very productive, we work well as a team, and the quality of our work is very good} \cite{bahli2005group}. Table~\ref{table:teamPerformance} classifies the two approaches.

\begin{table}
\caption{Approaches for Measuring Software Team Performance. Taken From \cite{purna2011soft}}
\centering
\begin{tabular}{ | p{4.0cm} | p{4.0cm} |} \hline
	M1 - Objective measures \begin{itemize}
		\item Function Points
		\item KLOC
		\item Object Points
		\item Use Case Points
		\item Defect Rates
		\item Defect Density
		\item Quantitative Metric
	\end{itemize}  &
	M2 - Subjective\slash Perceptual Team performance Ratings By:
		\begin{itemize}
		\item Team Members
		\item Management
		\item Customer
	\end{itemize} \\ \hline
	
\end{tabular}

\label{table:teamPerformance}
\end{table}

Similarly, \cite{ramasubbu2007globally} concluded in another study that software teams performance is measured in terms of function points per person hour and conformance to quality. The conformance quality refers to the defect rate claimed by the customer during acceptance testing. Team's adherence to budget and schedule is another measure of performance reported by \cite{boehm1981software}. According to \cite{purna2011soft}, a team's performance is a function of what individual team members are doing. More specifically, a successful team is one that is characterized by the following: 1) shared leadership roles, 2) specific and clear goals, 3) mutual accountability, 4) collective problem solving.

Albero Pomar et al. \cite{albero2014understanding} proposed two techniques for predicting future performance of scrum software teams. The first approach relies on plotting the accrued velocity for all previous sprints in order to identify the trends (downward or upward) of performance. The second approach depends on calculating a confidence interval to comprehend the probability of future velocities. They also proposed exploiting a traditional (non-agile) project management metric to gauge the amount of completed work over the planned work. The metric is calculated as the ratio of the total earned points over the total points planned in the sprint planning meeting \cite{albero2014understanding}. 

\[Schedule Performance Indicator
  = \dfrac{Earned Points}{Planned Points} * 100
\]

\section{Method}\label{ch:research_methodology}
The objective of this paper is to investigate and analyze whether group maturity is related to aspects of the performance of software development teams. More specifically, performance is examined by measuring both planning effectiveness and development velocity of four participating work groups from company A.
\paragraph{Research Questions}

This study aims to contribute to answering of the following questions.

\begin{enumerate}
	
	\item What is the association between group maturity and planning effectiveness?
	
    \item What is the association between group maturity and software development velocity? 

\end{enumerate}
  	
Group maturity in the four participating work groups was measured using the GDQ. The software development velocity was in turn measured by calculating the number of hours spent on developing scrum tasks for each member in the participating teams whereas planning effectiveness was assessed by using the Schedule Performance Indicator metric. 

\subsection{Case}

A combination of qualitative and quantitative data was used in this study. According to \cite{methodologyguidelines}, a case study is a suitable methodology for software engineering research, since it provides a deeper understanding of the phenomena under study. As a result, a case study was selected as the most suitable means for conducting this research. Using both qualitative and quantitative data provides an in-depth understanding to the way the participating groups are functioning and facilitates a better comparison between the groups.
\subsection{Subject Selection}

\paragraph{Company Description}
Company A is a Swedish company with 1,400 employees located in four different countries. The company is active in the fields of software development and business development. The increasing growth of the company's market share has stemmed a need for the company to work towards achieving more efficient and effective ways to develop its products. Part of their development effort is spent on developing the group dynamics in their software development teams. This research was conducted in collaboration with the company's staff at their branch located in Gothenburg, Sweden.

\paragraph{Work Groups in Company A}

First, we would like to reinforce the distinction, made in introduction section, between teams and work groups for the purpose of clarifying the terms we used in this research. A team is a structured group of individuals who share well-defined common goals that require coordinated interactions in order to effectively accomplish their tasks. A work group, on the other hand, is one in which members accomplish their tasks successfully, but not necessarily coordinate well and share the same goals \cite{teamsgroups}. Accordingly, we decided to use the term \emph{work groups} to refer to the participating groups in this research. Additionally, we gave anonymous names to the work groups to keep their identity unknown.

 Four software development work groups adopting scrum participated in this research. All work groups were formed eight to 40 months prior to the date of conduct of this research.\ Groups' size comprised of three to six members with ages ranging from 20 to 60 years. The duration of which the work groups were formed and had practiced scrum ranged between eight to 40 weeks. All work groups are cross-functional, which means that the skill-sets of members within each work group were homogeneously distributed and that members have the necessary skills to perform multiple essential roles in the development process. All of the work groups receive work packages, analyzed and defined by company B, which acts as the main customer for company A. These work packages shape out the work groups' product backlogs, which contain a number of requirements, written in the form of user stories, from which teams select and plan their development cycles (or sprints) respectively. The assignment of work packages to the work groups is done based on the their competence level.    

\subsection{Data Collection} \label{datacollection}
Estimations of user stories are done using planning poker, which is used to estimate the complexity in unit of points for either new features or change requests. Each work group collaborates closely with a designated product owner assigned by company B to represent the business, prioritize requirements, and conveys the product vision. It is important to mention that the selection of user stories, done at every planning and review meeting, is based on the priority of requirements conveyed by the product owner to the development group rather than being based on members' preference. Participating groups use a web-based project management and issue tracking tool. This allows them to manage their projects and visualize their work progress at any point in time. Stories are located at the leftmost part of the UI and are moved to the right as stories progress towards completion. This UI is divided into seven columns, starting from the far left: new, in progress, needs review, blocked, closed, and rejected. The column in progress indicates scrum tasks that have been assigned to an individual for development. The column \emph{blocked} contains all the stories that are temporary blocked because of other external dependencies or the absence of the assignee. On the other hand, column \emph{closed} refers to the  stories that were completed by members. 

The selection of software development work groups was carried out with help from a gatekeeper at company A. Data were collected from the work groups (N=4) at their work site during regularly scheduled meetings, with all members of each respective work group present. We used multiple data sources to increase the validity of the findings. Below are the data collection steps arranged in chronological order.

\paragraph{Unstructured Interviews}
Brief interviews of approximately 15 minutes each with the scrum master of each work group were conducted at the onset of the data collection process. These interviews allowed the author to gain a better understanding of the context of the groups' work and to schedule for the GDQ fill-out sessions and semi-structured interviews with the 19 participants from the four work groups. Some scrum masters were interviewed twice over the course of this study as new issues emerged.

\paragraph{The Maturity Levels of the Groups}
To examine the maturity level of the participating groups, the GDQ was used to obtain the members' perception about how each group is functioning. Individuals were requested to answer the sixty questions of the GDQ. All the GDQ fill-out sessions occurred during the last week of the group's ongoing sprints. This time was chosen to give the work groups the longest time possible in the sprint to resolve any issue related to their dynamics. A background variable in the GDQ is a question regarding the perceived productivity of the work group rated from not productive at all to highly productive.  

\paragraph{Development Velocity} 
In this study, the velocity of the four participating work groups was measured by calculating the mean of hours spent on implementing a number of new scrum tasks that were planned as part of new features. Scrum tasks were chosen over user stories since tasks, unlike stories, share similar complexity as each corresponds to a small unit of work planned by the scrum team \cite{tasks}. This method of measurement was discussed and approved by Company A. 
To measure the velocity of the work groups, access to their task boards was granted and data about velocity was collected at the end of the same sprint when the GDQs were administered. For each work group, an average of 40 completed (closed) tasks, planned under new features, were arbitrary selected from their last development cycle whereby eight tasks on average were taken per individual. Consequently, the difference between the end and start time (in unit of hours) for each task was computed and deducted from the total time in which the task was \emph{blocked}. A given task may get blocked in the event of a disruption caused by an external dependency or an unexpected member drop-out/absence. For example, if the assignee was on a leave, the status of his \emph{in progress} tasks will be temporary set to blocked. This requires the assignee to remember changing the status of the task to \emph{blocked} before he leaves and back to \emph{in progress} once he comes back. The scrum masters of the four work groups were requested, prior to the start of the sprint, to inform their members to ensure updating their tasks status promptly with every change. This will mitigate the risk of encountering skewness in the data resulting from members forgetting to update the status of tasks.

 Subsequently, the mean value of tasks accomplishment, for each work group, was calculated and recorded as the group's velocity. 
 \[Mean Velocity
  = \frac{\sum_{i=1}^{i=n}((End time - Start time)-Blocked time)}{n}
\] 
Ultimately, the first author sent the computed velocity with the IDs of the selected tasks to the scrum master of each work group to perform cross-checking on the computed velocity. Two scrum masters reported some errors in the computation of the mean values as they pinpointed a lack of compliance of two members in updating the status of five tasks respectively. Consequently, we recomputed and recorded the mean velocity of the two work groups.     
\paragraph{Planning Effectiveness}
Since all of the four participating work groups adopt scrum as their development methodology, they decide what can be accomplished in each sprint during their planning and review meetings.\ Accordingly, teams take into consideration the complexity of stories, the group's availability, and their technical competence level in planning what they can commit to in each sprint.\
The planning effectiveness of the work groups was measured using the Schedule Performance Indicator metric, which calculates the ratio of their total earned points over the total planned points for a given sprint.

\[ Schedule Performance Indicator
  = \dfrac{Earned Points}{Planned Points} * 100
\]
The mean planning effectiveness for all the sprints, which were selected according to two criteria, for each work group was calculated. The first criterion for the sprint selection is that the structure of the work groups remained unchanged, that is, no individuals joined or left the work group. The second is that their maturity level remained stable. This was confirmed by the interviewed group members during the semi-structured interviews when participants were asked \say{How long has the team's maturity level been stable?} allowing for an estimation of the duration that their work group maturity has not changed. Table~\ref{tab:planning} shows the total number of planned versus earned story points for each selected sprint for all the participating work groups. As can be seen, the number of sprints from which the planned and earned points were collected varied considerably. This reflects the difference in the duration that a given group's maturity remained unchanged. For example, the responses of the majority of members from group A during the semi-structured interview, revealed that their maturity has remained unchanged over the past ten sprints, in their opinion. Therefore, data from this period only was collected. On the other hand, the majority of members of work groups C and D agreed that their maturity has remained unchanged over the past four sprints. Therefore, the planned and earned points were collected from those sprints only. Table~\ref{tab:planning} also demonstrates considerable variations in the planned points of some sprints from groups C and D. For example, a drop in the planned points for the fourth sprints of group D was identified (from 40 to 9). These fluctuations in the planned points between sprints can be attributed to the cumulative experience attained over previous development iterations as well as to the availability of members within each respective group.  

\begin{table}
\caption{Planned vs. Earned Points}
\centering
    \begin{tabular}{|p{0.8cm}|p{1.0cm}p{0.2cm}p{0.2cm}p{0.2cm}p{0.2cm}p{0.2cm}p{0.2cm}p{0.2cm}p{0.2cm}p{0.2cm}p{0.3cm}|}
  
          \hline
           Group & \textbf{Sprint}& \textbf{1} & \textbf{2}& \textbf{3}& \textbf{4}  & \textbf{5}& \textbf{6}& \textbf{7}& \textbf{8} & \textbf{9} &   \textbf{10}  \\ \cline{1-12}
             \footnotesize  \centering A &  Planned  & \footnotesize4 & \footnotesize3   & \footnotesize5 & \footnotesize7 & \footnotesize4 & \footnotesize6 &        \footnotesize2 & \footnotesize4 & \footnotesize2 & \footnotesize 8\\ 
           &\footnotesize Earned & \footnotesize 0 & \footnotesize3 & \footnotesize3 &\footnotesize6 & \footnotesize2 & \footnotesize6 & \footnotesize2 & \footnotesize4 & \footnotesize2 & \footnotesize0\\ \hline 
           
            \footnotesize \centering B &\footnotesize Planned  & \footnotesize 4 & \footnotesize 4 & \footnotesize6 & \footnotesize8 & \footnotesize6 & \footnotesize5 & \footnotesize 2.5 & \footnotesize- & \footnotesize- & \footnotesize-\\ 
           &\footnotesize Earned &\footnotesize 2 & \footnotesize2 & \footnotesize 4 &\footnotesize 6 & \footnotesize0 & \footnotesize1 & \footnotesize0.5 &  \footnotesize- & \footnotesize- & \footnotesize-\\ \hline 
           
            \footnotesize \centering C &\footnotesize Planned  & \footnotesize22 & \footnotesize18 & \footnotesize14 & \footnotesize30 & \footnotesize- & \footnotesize- & \footnotesize- & \footnotesize- & \footnotesize- & \footnotesize-\\ 
            
           &\footnotesize Earned   & \footnotesize18 & \footnotesize10 &   \footnotesize11 & \footnotesize21 & \footnotesize- &           \footnotesize- & \footnotesize- & \footnotesize- &            \footnotesize- & \footnotesize-\\ \hline 
           
            \footnotesize \centering D &\footnotesize Planned  & \footnotesize80 & \footnotesize63 & \footnotesize40 & \footnotesize9 & \footnotesize- & \footnotesize- & \footnotesize- & \footnotesize- & \footnotesize- & -\\ 
           &\footnotesize Earned   & \footnotesize65 & \footnotesize24 &  \footnotesize21 & \footnotesize3 & \footnotesize- & \footnotesize- &  \footnotesize- & \footnotesize- & \footnotesize- & \footnotesize-\\ \hline

    \end{tabular}
    
    \label{tab:planning}
\end{table}

\paragraph{Semi-Structured Interviews}

A primary source of data collection was semi-structured interviews, a common way of interviewing in case study research \cite{qualitative}. These involve working from an interview guide -- a list of prepared questions and topics aimed at ensuring systematic and chronological coverage across interviews. However, the interview is flexibly conducted to allow for self-elaboration and exploration of emerging issues \cite{allison2007software}. In this research, the main purpose was to explore more issues of group development as well as to strengthen the validity of the responses obtained from the surveys (the GDQ). Following the interviewees' approvals to participate, an interview for each individual was taped, transcribed, and coded. 16 out of the 19 members agreed to have their interviews taped, while two members did not. As a result, the author did not include the latter as part of the data collection. 

\subsection{Data Analysis}
\paragraph{Normality Test}
A first step to decide which correlation method to use in the data analysis would be to evaluate if the data is normally distributed. We conducted a Shapiro-Wilk test for each residual value of the four GDQ scales and the velocity of the four participating work groups. The \emph{p} values for the velocity of groups A and C indicate statistical significance, where \emph{p}=.05 for group A and \emph{p}=.048 for group C. As a result, our normality assumption for our linear regression model is not valid. In addition, the Q-Q plot of residuals for velocity showed a wide scatter in the distribution of residuals across the regression line, which supports our finding from the Shapiro-Wilk analysis that our normality assumption is not valid. Spearman's rank-order correlation analysis was, therefore, selected as the most appropriate method to conduct the correlation for the collected data set.

\paragraph{Quantitative Data Analysis}
Spearman's rank-order correlation coefficient was used to investigate the connection between group maturity and development velocity, and between group maturity and planning effectiveness. Given the normality analysis check and the small sample size available in this research (four groups), Spearman's correlation was chosen as the most appropriate method to run the analysis, since it does not assume normality in the data. SPSS was used to aid in investigating the aforementioned correlations. For question one, Spearman's correlations were run on both individual and group level, using individual data (19 group members) and then using group data (four groups). For question two, Spearman's correlation was run on the group level only because the planning effectiveness is a group endeavour rather than an individual one. Running the analysis on the group and individual levels will reinforce the idea of the IMGD theory, which states that the dynamics of a particular group constitute the source of individual perceptions of that group. Moreover, it emphasizes the idea that groups, not individuals, should be the key element of any change efforts deemed important. Moreover, some group demographic background collected from the groups' responses on the GDQ were tested for correlation with the four group development scales. Specifically, this was done to examine the impact of the individuals' age, educational background, employment time in company A on the four different maturity scales.

\paragraph{Qualitative Data Analysis}
Thematic analysis was used to interpret the collected qualitative data. The data from the semi-structured interviews were collated into electronic documents, which made the process of handling, searching and comparing the large volumes of data more convenient and manageable. Data were broadly categorized into seven themes, which were related to the dynamics within the work groups, in order to address some of the issues in the four stages of group development (see Table~\ref{themes}). 

\begin{table}

  \caption{Themes Explored in Group Development} \label{themes} 
  
  \centering
  \begin{tabular}{cc}
  \hline
     \footnotesize \textbf{Themes} & \footnotesize \textbf{Stage} \\ \hline
     \footnotesize Leader Dependence & \footnotesize I \\
     \footnotesize Tentativeness and Politeness & \footnotesize I \\
     \footnotesize Participation and Cooperativeness & \footnotesize II \\
     \footnotesize Subgroups or Cliques & \footnotesize II \\
     \footnotesize Goal Clarity & \footnotesize III \\
     \footnotesize Structure & \footnotesize III \\
     \footnotesize Trust & \footnotesize III \\
     \footnotesize Goal Accomplishment & \footnotesize IV \\ \hline
  \end{tabular}
     
\end{table}  
 
Based on these themes, a list of seven questions was prepared to address issues related to the four GDQ scales. Additionally, the last question was asked to estimate the number of sprints to consider when calculating the planning effectiveness of each work group (see Table~\ref{table:semistructured}). The Nvivo software was used for transcribing and coding the data. 

\begin{table}
  \caption{Semi-structured Interview Questions}
  \centering
  
  \begin{tabular}{c}
  \toprule
       \footnotesize \textbf{Questions} \\ \hline
       \footnotesize What are your roles in the team? \\
       \footnotesize Are members overly polite to each other? \\
       \footnotesize Are members hesitant to ask for support from each other? \\ 
       \footnotesize Are there subgroups in the team? \\ 
       \footnotesize Is trust high in the team? \\
       \footnotesize Are you clear on your team goals?  \\
       \footnotesize What is causing delays in your sprint? \\
       \footnotesize How long has the team's maturity level been stable? \\ \hline
  \end{tabular}
  \label{table:semistructured}
\end{table}

\subsection{Ethical Considerations}

The importance of ethical standards of conduct for maintaining trust and collaboration with the participants in question has been highlighted by many authors \cite{runeson2009tutorial}. Participants were spoken to about the objectives of the research, the nature of their involvement, the measures that would be taken to protect their identity, and the right to not participate or to withdraw at any stage. This was first done during scheduled meetings with the scrum masters, then explained to the other group members during the first meeting. 

\section{Results}\label{sec:results}

\subsection{Maturity and Group Demography}

In order to examine the connection between some demographic information (age, years in company, and educational background) and group development, a Spearman's $\rho$ correlation analysis on the individual level was run. 
Overall, the results presented in table~\ref{table:descriptive_statistics} suggest that age relates to the group perceptions about ``trust and structure,'' i.e., the older the members were in a work group, the higher their perception about ``trust and structure'' gets. On the contrary, the number of employment years within a company is negatively correlated, on a moderate level, with the members' perception about their productivity. In other words, the more years the members spent in company A, the less productive they viewed their work groups. In this correlation analysis, the educational background played no role in the members' group development, according to the participants views.

\begin{longtable}{| p{.17\textwidth} | p{.12\textwidth} | p{.09\textwidth} | p{.09\textwidth} |p{.09\textwidth} | p{.09\textwidth} | p{.11\textwidth} |} \caption{Spearman's Correlations between Group Demography and Perception } \\ \hline 
		\label{table:descriptive_statistics}
		\footnotesize \textbf{Demography} & \footnotesize \textbf{Statistic} & \footnotesize \textbf{GDQ1} & \footnotesize \textbf{GDQ2} & \footnotesize \textbf{GDQ3} & \footnotesize \textbf{GDQ4} & \footnotesize \textbf{Productivity} \\ \hline
		\endhead 
		
		\begin{tabular}{c} \footnotesize Age \end{tabular}&
		\begin{tabular}{c} \footnotesize Coefficient \\ \footnotesize Sig. \\ \footnotesize N \end{tabular} & 
		\begin{tabular}{c} \footnotesize -0.352 \\ \footnotesize 0.139 \\ \footnotesize 19 \end{tabular} & 
		\begin{tabular}{c} \footnotesize 0.068 \\ \footnotesize 0.783 \\ \footnotesize 19  \end{tabular} & 
		\begin{tabular}{c} \footnotesize 0.455 \\ \footnotesize 0.050 \\ \footnotesize 19  \end{tabular} & 
		\begin{tabular}{c} \footnotesize 0.368 \\ \footnotesize 0.121 \\ \footnotesize 19  \end{tabular} & 
		\begin{tabular}{c} \footnotesize -0.248 \\ \footnotesize 0.305 \\ \footnotesize 19  \end{tabular} \\ \hline

		\begin{tabular}{c} \footnotesize Years In Company \end{tabular}&
		\begin{tabular}{c} \footnotesize Coefficient \\ \footnotesize Sig. \\ \footnotesize N \end{tabular} & 
		\begin{tabular}{c} \footnotesize -0.239\\ \footnotesize 0.325 \\ \footnotesize 19 \end{tabular} & 
		\begin{tabular}{c} \footnotesize 0.203 \\ \footnotesize 0.405 \\ \footnotesize 19  \end{tabular} &
		\begin{tabular}{c} \footnotesize 0.341 \\ \footnotesize 0.153 \\ \footnotesize 19  \end{tabular} & 
		\begin{tabular}{c} \footnotesize 0.220 \\ \footnotesize 0.366 \\ \footnotesize 19  \end{tabular} & 
		\begin{tabular}{c} \footnotesize -0.512 \\ \footnotesize 0.025 \\ \footnotesize 19  \end{tabular} \\ \hline 
		
		\begin{tabular}{c} \footnotesize Education \end{tabular}&
		\begin{tabular}{c} \footnotesize Coefficient \\ \footnotesize Sig. \\ \footnotesize N \end{tabular} & 
		\begin{tabular}{c} \footnotesize 0.315\\ \footnotesize 0.190 \\ \footnotesize 19 \end{tabular} & 
		\begin{tabular}{c} \footnotesize 0.039 \\ \footnotesize 0.872 \\ \footnotesize19  \end{tabular} &
		\begin{tabular}{c} \footnotesize 0.000* \\ \footnotesize 1.000 \\ \footnotesize 19  \end{tabular} & 
		\begin{tabular}{c} \footnotesize -0.102\\ \footnotesize 0.678 \\ \footnotesize 19  \end{tabular} & 
		\begin{tabular}{c} \footnotesize 0.090 \\ \footnotesize 0.715 \\ \footnotesize 19  \end{tabular} \\ \hline 	
\end{longtable}

\subsection{Maturity and Planning Effectiveness}
Since the sample size of our data set was small (N=4), a normality test was not performed on the residuals for the planning effectiveness.\ Therefore, Spearman's correlation was run to determine the connection between planning effectiveness and group development since it does not assume normality of the data. 

\paragraph{Correlation Analysis}
Since planning is a group endeavour, this correlation analysis was run on group level only. The results revealed a positive correlation between the fourth stage of group development and planning effectiveness and showed a significant convergent validity, i.e., the more mature a team is, the more effective they plan their sprints' stories thus deliver the expected outcome. While significant correlations were not found with scales I, II, and III, correlations on scale II and III are going in the right direction (see table \ref{table:corrPlan}). The correlation coefficient and significance (\emph{r} = 1 and \emph{p} = 0.000) describe the strength of the association between the two variables, the \emph{GDQ4} and \emph{Planning effectiveness}, which is a perfect positive one for this small data set. 

\begin{table}[h]
      \centering
       \caption{Correlations for GDQ Perceptions and Planning Effectiveness } 
      \begin{tabular}{| p{2.0cm} | p{1cm} | p{1cm} | p{1cm} | p{1cm}|}
       \hline
        \centering
         \footnotesize \textbf{Scale} &  \footnotesize \textbf{GDQ1} &  \footnotesize \textbf{GDQ2} &  \footnotesize \textbf{GDQ3} &  \footnotesize \textbf{GDQ4} \\ \hline
        
        \begin {tabular} {@{}l@{}} \small \textbf{Planning } \\ \footnotesize Sig. (2-tailed) \\ \footnotesize N
         \end{tabular} & 
        \begin {tabular} {@{}l@{}}  \footnotesize 0.400  \\ \footnotesize 0.6000 \\ \footnotesize4
        \end{tabular} & 
        
        \begin {tabular} {@{}l@{}}  \footnotesize -0.2000 \\ \footnotesize 0.8000 \\ \footnotesize 4
        \end{tabular} & 
        
        \begin {tabular} {@{}l@{}}  \footnotesize 0.4 \\ \footnotesize 0.6 \\ \footnotesize 4
        \end{tabular} &
        
        \begin {tabular} {@{}l@{}}  \footnotesize 1.000 \\ \footnotesize . \\ \footnotesize 4
        \end{tabular} \\ \hline
      \end{tabular}
      
      \label{table:corrPlan}
\end{table}

\paragraph{Planning Effectiveness Comparison}

Table~\ref{table:groupPlan} shows the planning effectiveness and the group development mean values of the four participating work groups. The evidence showed that work groups which scored higher in GDQ4 also scored higher in planning effectiveness. As can be seen from the table, Group D scored the highest GDQ4 score, compared to the other work groups, with a mean value of 64.67. It also outperformed the other work groups in planning effectiveness with a mean value of 71.48. On the other hand, the lowest GDQ4 mean value, 53.17, was scored by work group B, which exhibited the minimum planning effectiveness with a mean value of 40.2. 

\begin{table}
      \centering
       \caption{ Plannng Effectiveness and Group Development Mean values} 
      \begin{tabular}{| p{1.0cm} | p{1.4cm} | p{0.9cm} | p{0.9cm} | p{0.9cm}| p{0.9cm}|}
       \hline
        \centering
         \footnotesize \textbf{Group} &  \footnotesize \textbf{Planning} & 
         \footnotesize \textbf{GDQ1} &  \footnotesize \textbf{GDQ2} &  \footnotesize \textbf{GDQ3} &  \footnotesize \textbf{GDQ4} \\ \hline
        
        \begin {tabular} {@{}c@{}} \footnotesize  A \\ \footnotesize  B \\ \footnotesize  C \\ \footnotesize  D
         \end{tabular} & 
        \begin {tabular} {@{}c@{}}  \footnotesize 66.11  \\ \footnotesize 40.2 \\ \footnotesize 51.27 \\ \footnotesize 71.48
        \end{tabular} & 
        
        \begin {tabular} {@{}c@{}}  \footnotesize 40.80 \\  \footnotesize 40.33 \\  \footnotesize 44 \\  \footnotesize 42
        \end{tabular} & 
        
        \begin {tabular} {@{}c@{}} \footnotesize 31.60 \\ \footnotesize 37.67 \\\footnotesize 29.4 \\\footnotesize 32
        \end{tabular} &
        
        \begin {tabular} {@{}c@{}} \footnotesize 62.60 \\ \footnotesize 54.67 \\\footnotesize 56.8 \\\footnotesize 55.67 
        \end{tabular} &
        
        \begin {tabular} {@{}c@{}} \footnotesize 63.80 \\ \footnotesize 53.17 \\\footnotesize 60.20 \\\footnotesize 64.67 
       \end{tabular} \\ \hline
       
      \end{tabular}
      
      \label{table:groupPlan}
\end{table}

\begin{figure*}
  \centering
  {\includegraphics[width=0.50 \textwidth]{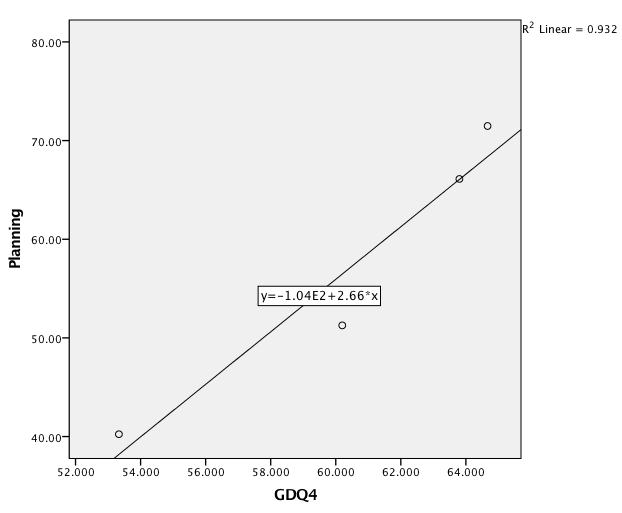}\label{fig:scatter:GDQ4}}
  \caption{Planning Effectiveness and Group Development Mean Values}
   \label{fig:scatter:Planning_GDQ}
\end{figure*}

\Cref{fig:scatter:Planning_GDQ} shows the scatter plot for the planning effectiveness as the dependent variable and the fourth group development scale (GDQ IV) as the independent variable. Each dot represents one of the four participating groups with the \emph{x} coordinate as the group development mean and the \emph{y} coordinate as the planning effectiveness mean value. Figure \ref{fig:scatter:Planning_GDQ} shows that $R^2$ = 0.93, which means that 93.2\% of the variance in the planning effectiveness can be explained by the fourth scale of the group development (GDQ4). This conclusion is built on the assumption that our population data is linear.

\subsection{Maturity and Development Velocity}
The analysis performed on a group and an individual level helped reinforce 
the notion that dynamics of a particular group constitute the source of individual perceptions of that group. Tables~\ref{table:groupVel} and~\ref{table:groupInd} clearly demonstrate that the findings are the same, i.e., no correlation exists between development velocity and maturity on both the individual and group levels.

\paragraph{Correlation Analysis}

  A second Spearman's correlation was conducted to determine the association between the work groups' perception about their maturity, on all the GDQ scales, and their development velocity. On both the group and individual levels, no significant relationship was identified (see \cref{table:groupVel,table:groupInd}). 
\begin{table}
      \centering
       \caption{Correlations between GDQ scales and Velocity -- Group Level} 
      \begin{tabular}{| p{2.0cm} | p{1cm} | p{1cm} | p{1cm} | p{1cm}|}
       \hline
        \centering
        \small \textbf{Scale} & \footnotesize \textbf{GDQ1} & \footnotesize \textbf{GDQ2} & \footnotesize \textbf{GDQ3} & \footnotesize \textbf{GDQ4} \\ \hline
        
        \begin {tabular} {@{}l@{}} \textbf{Velocity} \\ \footnotesize Sig. (2-tailed) \\  \footnotesize N
         \end{tabular} & 
        \begin {tabular} {@{}l@{}} \footnotesize 0.2000  \\ \footnotesize 0.8000 \\ \footnotesize 4
        \end{tabular} & 
        
        \begin {tabular} {@{}l@{}} \footnotesize -0.4000 \\ \footnotesize 0.6000 \\ \footnotesize 4
        \end{tabular} & 
        
        \begin {tabular} {@{}l@{}}  \footnotesize 0.8000 \\ \footnotesize 0.2000 \\ \footnotesize 4
        \end{tabular} &
        
        \begin {tabular} {@{}l@{}} \footnotesize 0.800 \\ \footnotesize 0.2000 \\ \footnotesize 4
        \end{tabular} \\ \hline
      \end{tabular}
      
      \label{table:groupVel}
\end{table}

\begin{table}
      \centering
       \caption{Correlations between GDQ scales and Velocity -- Individual Level} 
      \begin{tabular}{| p{2.0cm} | p{1cm} | p{1cm} | p{1cm} | p{1cm}|}
       \hline
        \centering
        \small \textbf{Scale} & \footnotesize \textbf{GDQ1} & \footnotesize \textbf{GDQ2} & \footnotesize \textbf{GDQ3} & \footnotesize \textbf{GDQ4} \\ \hline
        
        \begin {tabular} {@{}l@{}}  \textbf{Velocity} \\ \footnotesize Sig. (2-tailed) \\ \footnotesize N
         \end{tabular} & 
        \begin {tabular} {@{}l@{}} \footnotesize 0.310  \\ \footnotesize 0.196 \\ \footnotesize 19
        \end{tabular} & 
        
        \begin {tabular} {@{}l@{}} \footnotesize -2.16 \\ \footnotesize 0.374 \\ \footnotesize 19
        \end{tabular} & 
        
        \begin {tabular} {@{}l@{}} \footnotesize 0.236 \\ \footnotesize 0.330 \\ \footnotesize 19
        \end{tabular} &
        
        \begin {tabular} {@{}l@{}} \footnotesize 0.204 \\ \footnotesize 0.402 \\ \footnotesize 19
        \end{tabular} \\ \hline
      \end{tabular}
      
      \label{table:groupInd}
\end{table}
\paragraph{Velocity Comparison}
Table~\ref{table:descriptive_statistics_velocity} shows the velocity and the group development mean values of the four participating work groups, which were both measured during the same sprint. As can be seen from the table, work group B has the minimum velocity mean value of 27.33, while work group A scored the highest mean with a value of 48.07. As shown in table~\ref{table:groupVel}, no connection between the groups' mean velocity and the group development scales was found. 

\begin{table}
      \centering
       \caption{Velocity Mean Values and Group Development Mean Values} 
      \begin{tabular}{| p{.9cm} | p{1.2cm} | p{0.9cm} | p{0.9cm} | p{0.9cm}| p{0.9cm}|}
       \hline
        \centering
         \footnotesize \textbf{Group} &  \footnotesize \textbf{Velocity} &  \footnotesize \textbf{GDQ1} &  \footnotesize \textbf{GDQ2} &  \footnotesize \textbf{GDQ3} &  \footnotesize \textbf{GDQ4} \\ \hline
        
        \begin {tabular} {@{}c@{}} \footnotesize  A \\ \footnotesize  B \\ \footnotesize  C \\ \footnotesize  D
         \end{tabular} & 
        \begin {tabular} {@{}c@{}}  \footnotesize 48.07  \\ \footnotesize 27.33 \\ \footnotesize 35.29 \\ \footnotesize 47.82
        \end{tabular} & 
        
        \begin {tabular} {@{}c@{}}  \footnotesize 40.80 \\  \footnotesize 40.33 \\  \footnotesize 44 \\  \footnotesize 42
        \end{tabular} & 
        
        \begin {tabular} {@{}c@{}} \footnotesize 31.60 \\ \footnotesize 37.67 \\\footnotesize 29.4 \\\footnotesize 32
        \end{tabular} &
        
        \begin {tabular} {@{}c@{}} \footnotesize 62.60 \\ \footnotesize 54.67 \\\footnotesize 56.8 \\\footnotesize 55.67 
        \end{tabular} &
        
        \begin {tabular} {@{}c@{}} \footnotesize 63.80 \\ \footnotesize 53.17 \\\footnotesize 60.20 \\\footnotesize 64.67 
       \end{tabular} \\ \hline
       
      \end{tabular}
      
      \label{table:descriptive_statistics_velocity}
\end{table}

\subsection{Semi-Structured Interviews} \label{semi_structure}

Responses from interviewees were thematically analyzed. Below are the main results of this analysis. 

\paragraph{Roles in The Work Groups}
All members from the participating work groups were able to describe their roles with ease, which means they were all clear on their responsibilities. 
 
\paragraph{Politeness and Tentativeness}
The majority of members in work group A (75\%) did not consider over politeness evident in their work group, whereas 25\% considered politeness to be a \emph{rare occasion} in the work group. Members of work group B explained that over politeness depends on the situation and the member's personality rather than a general trait for the work group. All of the interviewed members from work group C agreed that members were overly polite with each other. 75\% of them linked this over politeness to the nature of engineers and their cultural backgrounds whereas 25\% said that members exhibited over politeness only in process related situations rather than technical ones.\ Finally, members from work group D perceived each other as being overly polite and associated this to the fact that they are newly formed and have not yet built a cohesive relationship with each other. The exception to this is one work group member who has been in the team for the longest time, perceived the work group not extremely polite. 
  
\paragraph{Cooperativeness and Support}
All of the interviewees from work group A reported that they are not hesitant to reach out each other when needed. 50\% of members from work group B suggested that people are not hesitant to ask for support from any one in the work group. One of those linked this to the fact that members do not want to take responsibilities on their behalf \emph{they might ask anyone so that they do not take the responsibility}. On the other hand, the remaining 50\% of members suggested that members are sometimes hesitant and linked this to the personality and topic type. 25\% of the interviewees from work group C shared a consensus that members are hesitant to seek for support from each other \emph{People are very concerned with each other and are reluctant to ask for support} whereas 75\% suggested that members tend to seek support from knowledgeable people in the work group, regardless of who they are. Finally, the majority of members in work group D explained that members are a bit hesitant to seek for support and explained that it is related to the fact that they have not yet had enough time to interact with each other and create a cohesive relationship. On the contrary, one member explained that members reach each other and are not scared to point out problems. 
 
\paragraph{Subgroups and Cliques} 
75\% of the interviewed members from work group A believed that there are no sub-work groups whereas 25\% agreed that there are subgroups. work group B explained that their work group is divided into two subgroups, over half of those linked this to the \emph{age range} factor \emph{there are two subgroups, and they are about the same age and stage in life}. All of the interviewed members from work group C explained that people with more technical knowledge or similar interests formed cliques in this work group. In addition, a third of the members in work group D believed that there are no cliques in their work group, the second third suggested the occurrence of subgroups, and the last third were not able to tell.

\paragraph{Trust} 
All interviewed members from work group A and C believed that trust is high in their work groups. Opinions in work group B is divided, whereby 40\% believed there is no trust among members, 40\% believed that there is trust, while the remaining mentioned that trust is a relative issue in the work group. Members in work group D distinguished between internal and external trust. They referred to external trust as the way in which external teams in company A perceive their work group; whereas internal trust was defined as how much members in the work group trust each other. All members in work group D believed that the external trust is high: \emph{From outside the team, trust is pretty solid. If something should be done, people trust us}. On the other hand, a third of the members explained that trust is high but not satisfactory yet: \emph{Maybe in the long run we will probably build higher internal trust}, while two thirds of the members suggested a lack of internal trust (within the work group) \emph{I don't have trust nor members have it to others}. 
 
\paragraph{Goal Clarity}
75\% of members in work group A mentioned some of their goals on a sprint level only: \emph{we only look at the sprint goals} whereas 25\% of them did not know any goals, whether on a sprint level or not. The majority of members from work group B could not mention any long or short term goals for their work group \emph{There are no common team goals, but rather some individual goals to reach}. 50\% of them linked this to the poorly defined customer specifications while almost 20\% believed that members are too focused on the development that they forget the work group goals: \emph{Most of them would remember them if you remind them but not everyone realizes that they know them}. On the contrary, all members from work group C recited their short term goals and not the long term ones. There was a 50\% overlap in their answers. Finally, third of the members from work group D were able to recite some of their work group's goals, while two third of the individuals could not.
 
\paragraph{Delays in Goal Accomplishment } 
The majority of members of work group A agreed that their lack of knowledge in one particular software engineering discipline (kept anonymous here) is negatively affecting their commitment to achieving their goals. Half of the members from work group B explained that the main reason for their delay was the unclarity of requirements received from their customer company: \emph{We don't know what we are doing. We need clear requirements}. The remaining 50\% had a common view that the external dependencies, the underestimation of workload, and the lack of knowledge in the domain of work are the reasons for their delay. 25\% of members from work group C attributed the delay to not knowing how to work well enough as a new team, whereas 75\% of them gave a common explanation, which persisted in their lack of experience to estimate the time needed for code review. 25\% of those gave additional reasons such as external dependencies and sick leaves. The other 25\% suggested that delays were the result of lack of knowledge in coding and the product, and underestimation of refactoring. Finally, member of work group D had extremely different explanations for their delay, which mainly persists in their lack of knowledge in the product and the lack of norms within the work group, and the variations in the level of technical competencies within the work group.  
 
\paragraph{Stability of Maturity in The Work Groups} 
Finally, all of the interviewees from work group A presumed that their maturity level has not changed during the last ten months. The majority of members from work group B suggested that their maturity has been stable for six months. 75\% of members in work group C believed that their maturity level has been stable for three months, whereas 25\% could not proximate a specific period. Members from work group D expressed different views about the period of stability. 75\% agreed that their group maturity remained unchanged for two months, whereas 25\% suggested that the group maturity has continuously progressed and it stopped progressing one month ago. 

The qualitative analysis revealed that work groups B and D experienced the highest number of group development issues explored in the semi-structured interviews whereas work groups A and C showed the lowest number of these issues. Also, a disparity of viewpoints was evident from the opinions of members in work groups B and D where individuals perceived their work groups' to be functioning differently. The major issues that emerged in work groups B and D seemed to relate to the different technical knowledge or age range of members. Some members with high technical knowledge (more experienced) tend to prefer working collaboratively with each other rather than working with individuals who had less technical experience. This may explain the lack of trust and goals' clarity between individuals in both group B and D.  

\section{Discussion}\label{sec:disc}

\subsection{Reflection on Efficiency and Effectiveness}
 We emphasize on interpreting the results in light of the distinction between efficiency and effectiveness. Our velocity measurement only reflects the efficiency of work groups in accomplishing scrum tasks, with no indication on how effective they were implemented. On the other hand, the measurements of planning effectiveness reveals the work groups' ability to deliver the expected outcome within the planned time frame.

\subsection{Answers to Research Questions}

\paragraph{RQ1 - What is the association between group maturity and planning effectiveness?}
In this research, we investigated the relationship between four independent variables (the group development stages) and the planning effectiveness. The results showed a perfect positive correlation (+1.0) between the fourth GDQ scale and the planning effectiveness among the four participating work groups, which means that both variables move in a strong tandem with each other and are positive in 100\% of the time. In other words, the higher a software development team scores on the measurement of the fourth group development phase, the more effective it becomes in planning its requirements. This supports the findings of other studies which confirm that task performance and work activity occur at higher levels later in a group's development~\cite{school,intensivecare,facilitating}. The significance of this research is that it provided evidence to support a relationship between group development and team performance in software engineering context. 

The overall conclusion drawn from the qualitative analysis overlapped with those revealed from the quantitative ones, such that they provided further evidence for the validity of the interviewees' responses to the GDQ with respect to the topics explored in the semi-structured interviews. For example the thematic analysis revealed that members from work group B had the highest number of group development issues, compared to the other work groups. Contrary to work group B, members of work group A had the lowest number of issues, which might be an indicator that this former work group is at the higher levels of group development.

\paragraph{RQ2 - What is the association between group maturity and software development velocity?}

We investigated the connection between the two variables, group development and velocity. The motivation for investigating this research question was to address a gap recited by \cite{maturityAgility} in which a positive correlation between maturity and velocity would support their findings about the connection between agility and group maturity. The results drawn from our analysis to this question were not in concordance with what \cite{maturityAgility} suggested, since we could not provide an empirical evidence to support a significant convergent validity between group maturity and development velocity. The analysis of the qualitative data revealed that the majority of participants linked their tasks development delays to technical and process related aspects rather than issues pertinent to the dynamics and norms within their work groups. This shows an interesting and non-predicted dependence on technical skills and process-related aspects in the software engineering domain that might be different compared to performance aspects in other fields. 

\subsection{Implications for Research and Practice}

We will now present some of the possible improvements that would increase the software development team performance. The first would be to motivate software developers to focus more on discussing and clarifying their work group goals. By this we mean that members should work more on achieving their group goals rather than focusing on the individual ones only. Our research suggests that the more effective work groups know their group goals, which is in alignment with stage III of the IMGD model which suggests that clarity of goals contribute to the development of more productive and work-focused groups. The second would be to motivate software developers to freely discuss and communicate process-related issues rather than only discussing the technical ones as members respectively reported a tendency of hesitance to ask for support in process-related issues. The third would be to consider having team members of diverse backgrounds working together in order to allow building more trust and structure within teams. This is supported by research done by \cite{roberge} who attempted to address when and how diversity in teams leads to better performance by conceptualizing a multi-level model that identifies the psychological mechanisms that explain how diversity can have a positive impact on the performance of teams. On the group level, these psychological mechanisms were identified as communication, group involvement, and group trust \cite{roberge}. Although our research only included one aspect of diversity, which is age, our qualitative and quantitative analysis clearly show that the age of software development team members relates to their perceptions of \emph{trust and structure}. In addition, work groups need to be given the opportunity to mature over time in order to achieve higher planning effectiveness; thus, becoming a self-organizing unit where all members can provide input for accurate sprint planning.  

\subsection{Validity Threats}
 
One needs to be careful when generalizing the findings of this study outside this specific case because only four participating groups from the same company were studied, which is a small sample. However, the combination of the data collection methods we used in this research, qualitative interviews and quantitative surveys would triangulate our findings, thus would strengthen the validity of our results.
 
This area of research is sensitive to the participating members since it involves the disclosure of the dynamics within their groups to us, which may influence the validity of the groups' responses to the quantitative survey and the qualitative interviews. At the onset of this research, an attempt to mitigate this was made by explaining the research purpose to the participants and by confirming the anonymity of their responses. However, it is not possible to refute the presence of bias in the participants' responses on both the surveys and the interviews. To reinforce the anonymity of the participating work groups, we avoided stating any information that would indicate their identity. Moreover, a self bias in the coding process of the semi-structured interviews cannot be ruled out and requires a second coder to validate the responses and thus minimize the self bias.
Our measurement of velocity relied on measuring the time spent by each work group on tasks accomplishment, which means that the amount of teamwork required to accomplish those tasks may not be significant, and an individual endeavor on each task may be sufficient to get the work done. This may provide an explanation to the absence of correlation between velocity and group maturity. Finally, although the majority of members explained that they instantly close their tasks after finishing their implementations, we can not guarantee that all of the tasks we selected for analysis were closed this way. This may have had an effect on the validity of our velocity measurement.  

\section{Conclusion and Future Work}
In the course of this research, we aimed at investigating how development velocity and planning effectiveness of software development teams relate to the four phases of their group development. Our findings showed that the fourth stage of group development, in the adapted framework, is significantly related to the effectiveness of software teams in planning their requirements whereas no evidence was provided to conclude a similar relationship with development velocity. Moreover, it indicates that there are considerable differences as to how group development relates to the effectiveness and efficiency of software teams. That is, a team with a higher score on the measurement of group development is possibly a more effective one but not more efficient. We believe that this research provided additional knowledge to the prominence of the human interactions within software development teams. Particularly by providing empirical evidence about the link between group maturity and planning effectiveness. We believe that the knowledge provided is sufficient to trigger organizations to drive more focus on those aspects, as they may provide benefits to software development teams. 

We would like to see the results of similar studies conducted with larger sample sizes from different companies. Also, we would encourage further studies to expand more upon the connection between group development aspects and team performance in software development. For example by measuring, function points, defects rate, and kilo lines of codes to assess team performance. Finally, we would like to see the results of studies that combine several objective and subjective methods to assess the performance of software development teams and highlight how each relates to group maturity.  

\bibliographystyle{abbrv}
\bibliography{references}  

\begin{thebibliography}{10}

\bibitem{agarwal2012tracking}
M.~Agarwal and R.~Majumdar.
\newblock Tracking scrum projects: Tools, metrics and myths about agile.
\newblock {\em International Journal of Emerging Technology and Advanced
  Engineering}, 2:97--104, 2012.

\bibitem{albero2014understanding}
F.~Albero~Pomar, J.~A. Calvo-Manzano, E.~Caballero, and M.~Arcilla-Cobi{\'a}n.
\newblock Understanding sprint velocity fluctuations for improved project plans
  with scrum: a case study.
\newblock {\em Journal of Software: Evolution and Process}, 26(9):776--783,
  2014.

\bibitem{allison2007software}
I.~Allison and Y.~Merali.
\newblock Software process improvement as emergent change: A structurational
  analysis.
\newblock {\em Information and software technology}, 49(6):668--681, 2007.

\bibitem{anderson1998measuring}
N.~R. Anderson and M.~A. West.
\newblock Measuring climate for work group innovation: Development and
  validation of the team climate inventory.
\newblock {\em Journal of organizational behavior}, 19(3):235--258, 1998.

\bibitem{asproni2004motivation}
G.~Asproni.
\newblock Motivation, teamwork, and agile development.
\newblock {\em Agile Times}, 4(1):8--15, 2004.

\bibitem{bahli2005group}
B.~Bahli and M.~D. Buyukkurt.
\newblock Group performance in information systems project groups: An empirical
  study.
\newblock {\em Journal of Information Technology Education}, 4(1):97--113,
  2005.

\bibitem{rippleeffect}
S.~G. Barsade.
\newblock The ripple effect: Emotional contagion and its influence on group
  behavior.
\newblock {\em Administrative Science Quarterly}, 47(4):644--675, 2002.

\bibitem{boehm1981software}
B.~W. Boehm et~al.
\newblock {\em Software engineering economics}, volume 197.
\newblock Prentice-hall Englewood Cliffs (NJ), 1981.

\bibitem{facilitating}
G.~Buzaglo and S.~A. Wheelan.
\newblock Facilitating work team effectiveness: Case studies from central
  america.
\newblock {\em Small Group Research}, 30(1):108--129, 1999.

\bibitem{cissna1984}
K.~N. Cissna.
\newblock Phases in group development: The negative evidence.
\newblock {\em Small Group Research}, 15(1):3--32, 1984.

\bibitem{tasks}
M.~Cohn.
\newblock The difference between a story and a task.
\newblock
  https://www.mountaingoatsoftware.com/blog/the-difference-between-a-story-and-a-task,
  2015.
\newblock [Online; accessed: 11-April-2016].

\bibitem{verbal}
B.~L. Davidson.
\newblock {\em Verbal Behavior Patterns In Groups Of Different Ages}.
\newblock PhD thesis, Temple University, 2001.

\bibitem{evans1986}
N.~J. Evans and P.~A. Jarvis.
\newblock The group attitude scale a measure of attraction to group.
\newblock {\em Small Group Research}, 17(2):203--216, 1986.

\bibitem{integratingimprovement}
G.~A. Gack and K.~Robison.
\newblock Integrating improvement initiatives: Connecting six sigma for
  software, cmmi{\textregistered}, personal software process (psp)(sm), and
  team software process (tsp)(sm).
\newblock {\em Software Quality Profesional}, 5(13), 2003.

\bibitem{TSPRevolution}
G.~Goth.
\newblock In the news.
\newblock {\em IEEE Software}, 17(6):125--127, Nov 2000.
\newblock Copyright - Copyright IEEE Computer Society Nov/Dec 2000; Last
  updated - 2014-05-19; CODEN - IESOEG.

\bibitem{gren2017links}
L.~Gren.
\newblock The links between agile practices, interpersonal conflict, and
  perceived productivity.
\newblock In {\em Proceedings of the 21st International Conference on
  Evaluation and Assessment in Software Engineering}, pages 292--297. ACM,
  2017.

\bibitem{maturityAgility}
L.~Gren, R.~Torkar, and R.~Feldt.
\newblock Group maturity and agility, are they connected? -- a survey study.
\newblock In {\em 41st EUROMICRO Conference on Software Engineering and
  Advanced Applications (SEAA)}, pages 1--8. IEEE, 2015.

\bibitem{grenjss}
L.~Gren, R.~Torkar, and R.~Feldt.
\newblock The prospects of a quantitative measurement of agility: {A}
  validation study on an agile maturity model.
\newblock {\em The Journal of Systems and Software}, 107:38—--49, 2015.

\bibitem{grenjss2}
L.~Gren, R.~Torkar, and R.~Feldt.
\newblock Group development and group maturity when building agile teams: {A}
  qualitative and quantitative investigation at eight large companies.
\newblock {\em The Journal of Systems and Software}, 124:104—--119, 2017.

\bibitem{TSPinGSD}
A.~Hern{\'a}ndez-L{\'o}pez, R.~Colomo-Palacios, {\'A}.~Garc{\'\i}a-Crespo, and
  P.~Soto-Acosta.
\newblock Team software process in gsd teams: A study of new work practices and
  models.
\newblock {\em International Journal of Human Capital and Information
  Technology Professionals (IJHCITP)}, 1(3):32--53, 2010.

\bibitem{hinsz1997emerging}
V.~B. Hinsz, R.~S. Tindale, and D.~A. Vollrath.
\newblock The emerging conceptualization of groups as information processors.
\newblock {\em Psychological bulletin}, 121(1):43--64, 1997.

\bibitem{hoegl2001teamwork}
M.~Hoegl and H.~G. Gemuenden.
\newblock Teamwork quality and the success of innovative projects: A
  theoretical concept and empirical evidence.
\newblock {\em Organization science}, 12(4):435--449, 2001.

\bibitem{huckman2009team}
R.~S. Huckman, B.~R. Staats, and D.~M. Upton.
\newblock Team familiarity, role experience, and performance: Evidence from
  indian software services.
\newblock {\em Management science}, 55(1):85--100, 2009.

\bibitem{humphrey2000guest}
W.~S. Humphrey.
\newblock Guest editor's introduction: The personal software process-status and
  trends.
\newblock {\em IEEE Software}, 17(6):71, 2000.

\bibitem{teamsgroups}
J.~Karn, S.~Syed-Abdullah, A.~J. Cowling, and M.~Holcombe.
\newblock A study into the effects of personality type and methodology on
  cohesion in software engineering teams.
\newblock {\em Behaviour \& Information Technology}, 26(2):99--111, 2007.

\bibitem{grupp}
J.~Keyton.
\newblock {\em Communicating in groups: Building relationships for group
  effectiveness}.
\newblock McGraw-Hill, New York, 2002.

\bibitem{liang2007effect}
T.-P. Liang, C.-C. Liu, T.-M. Lin, and B.~Lin.
\newblock Effect of team diversity on software project performance.
\newblock {\em Industrial Management \& Data Systems}, 107(5):636--653, 2007.

\bibitem{qualitative}
L.~McLeod, S.~G. MacDonell, and B.~Doolin.
\newblock Qualitative research on software development: a longitudinal case
  study methodology.
\newblock {\em Empirical Software Engineering}, 16(4):430--459, 2011.

\bibitem{mills}
T.~M. Mills.
\newblock {\em The sociology of small groups}.
\newblock Prentice-Hall Englewood Cliffs, NJ, 1967.

\bibitem{ong2005team}
A.~Ong, G.~W. Tan, and A.~Kankanhalli.
\newblock Team expertise and performance in information systems development
  projects.
\newblock In {\em Proceedings of the 9th Asia Pacific Conference on Information
  Systems, Bangkok, Thailand, July}, pages 7--10, 2005.

\bibitem{purna2011soft}
G.~Purna~Sudhakar, A.~Farooq, and S.~Patnaik.
\newblock Soft factors affecting the performance of software development teams.
\newblock {\em Team Performance Management: An International Journal},
  17(3/4):187--205, 2011.

\bibitem{ramasubbu2007globally}
N.~Ramasubbu and R.~K. Balan.
\newblock Globally distributed software development project performance: An
  empirical analysis.
\newblock In {\em Proceedings of the the 6th joint meeting of the European
  software engineering conference and the ACM SIGSOFT symposium on the
  foundations of software engineering}, pages 125--134. ACM, 2007.

\bibitem{roberge}
M.~Roberge and R.~van Dick.
\newblock Recognizing the benefits of diversity: when and how does diversity
  increase group performance?
\newblock {\em Human Resource Management Review}, 20(4):295--308, 2010.

\bibitem{methodologyguidelines}
P.~Runeson and M.~H{\"o}st.
\newblock Guidelines for conducting and reporting case study research in
  software engineering.
\newblock {\em Empirical Software Engineering}, 14(2):131--164, 2009.

\bibitem{runeson2009tutorial}
P.~Runeson and M.~H{\"o}st.
\newblock Tutorial: Case studies in software engineering.
\newblock In {\em Product-Focused Software Process Improvement}, pages
  441--442. Springer, 2009.

\bibitem{sawyer2001effects}
S.~Sawyer.
\newblock Effects of intra-group conflict on packaged software development team
  performance.
\newblock {\em Information Systems Journal}, 11(2):155--178, 2001.

\bibitem{sidky}
A.~Sidky.
\newblock {\em A structured approach to adopting agile practices: The agile
  adoption framework}.
\newblock PhD thesis, Virginia Polytechnic Institute and State University,
  2007.

\bibitem{humphrey2000personal}
R.~Sison.
\newblock Personal software process (psp) assistant.
\newblock In {\em 12th Asia-Pacific Software Engineering Conference (APSEC)},
  page~8. IEEE, 2005.

\bibitem{tuckman1965}
B.~W. Tuckman.
\newblock Developmental sequence in small groups.
\newblock {\em Psychological bulletin}, 63(6):384, 1965.

\bibitem{tuckman1977}
B.~W. Tuckman and M.~A.~C. Jensen.
\newblock Stages of small-group development revisited.
\newblock {\em Group \& Organization Management}, 2(4):419--427, 1977.

\bibitem{wheelan1994}
S.~Wheelan.
\newblock The group development questionnaire: A manual for professionals.
\newblock {\em Provincetown, MA: GDQ Associates}, 1994.

\bibitem{wheelan2005}
S.~A. Wheelan.
\newblock {\em Group processes: A developmental perspective}.
\newblock Allyn and Bacon, Boston, 2 edition, 2005.

\bibitem{intensivecare}
S.~A. Wheelan, C.~N. Burchill, and F.~Tilin.
\newblock The link between teamwork and patients' outcomes in intensive care
  units.
\newblock {\em American Journal of Critical Care}, 12(6):527--534, 2003.

\bibitem{wheelan1996}
S.~A. Wheelan and J.~M. Hochberger.
\newblock Assessing the functional level of rehabilitation teams and
  facilitating team development.
\newblock {\em Rehabilitation Nursing}, 21(2):75--81, 1996.

\bibitem{validationstudy}
S.~A. Wheelan and J.~M. Hochberger.
\newblock Validation studies of the group development questionnaire.
\newblock {\em Small Group Research}, 27(1):143--170, 1996.

\bibitem{school}
S.~A. Wheelan and F.~Tilin.
\newblock The relationship between faculty group development and school
  productivity.
\newblock {\em Small Group Research}, 30(1):59--81, 1999.

\end{thebibliography}

\end{document}